\documentclass{IEEEtran4PSCC}

\ifCLASSINFOpdf
   \usepackage[pdftex]{graphicx}
\else
   \usepackage[dvips]{graphicx}
\fi
\usepackage{lineno}
\usepackage{cite}
\usepackage{amsmath,amssymb,mathtools,underscore,subfigure}
\usepackage{bbold}
\usepackage[utf8]{inputenc}
\usepackage{url}
\usepackage{varioref}
\usepackage[colorlinks]{hyperref}
\usepackage[capitalize,nameinlink]{cleveref}
\AtBeginDocument{}

\DeclareFontFamily{U}{rcjhbltx}{}
\DeclareFontShape{U}{rcjhbltx}{m}{n}{<->rcjhbltx}{}
\DeclareSymbolFont{hebrewletters}{U}{rcjhbltx}{m}{n}
\DeclareMathSymbol{\shin}{\mathord}{hebrewletters}{152}

\newcommand{\abs}[1]{{{\left | {#1} \right |}}}
\newcommand{\skipall}[1]{}

\DeclareMathOperator*{\argmin}{arg\,min}

\let\emph\relax
\DeclareTextFontCommand{\emph}{\bfseries\em}

\DeclareMathOperator{\transp}{\negthinspace^\top}

\DeclareMathOperator{\sign}{sgn}

\newcommand{\eqdef}{\coloneqq}
\newcommand{\IR}{\ensuremath{\mathbb{R}}}
\newcommand{\IN}{\ensuremath{\mathbb{N}}}

\newcommand{\der}{\ensuremath{\text{d}}}

\newcommand{\dquot}[2]{\ensuremath{\frac{\der\thinspace{#1}}{\der{#2}}}}
\newcommand{\ddquot}[2]{\ensuremath{\frac{\der^2{#1}}{{#2}}}}

\DeclareMathOperator{\D}{D}
\DeclareMathOperator{\Hs}{H}

\newcommand{\ONE}{\mathbb{1}}

\renewcommand{\phi}{\varphi}
\renewcommand{\epsilon}{\varepsilon}

\newcommand{\funto}{\ensuremath \rightarrow}
\makeatletter
\let\old@ps@headings\ps@headings
\let\old@ps@IEEEtitlepagestyle\ps@IEEEtitlepagestyle
\def\psccfooter#1{%
    \def\ps@headings{%
        \old@ps@headings%
        \def\@oddfoot{\strut\hfill#1\hfill\strut}%
        \def\@evenfoot{\strut\hfill#1\hfill\strut}%
    }%
    \def\ps@IEEEtitlepagestyle{%
        \old@ps@IEEEtitlepagestyle%
        \def\@oddfoot{\strut\hfill#1\hfill\strut}%
        \def\@evenfoot{\strut\hfill#1\hfill\strut}%
    }%
    \ps@headings%
}
\makeatother

\psccfooter{%
        \parbox{\textwidth}{\hrulefill \\ \small{21st Power Systems Computation Conference} \hfill \begin{minipage}{0.2\textwidth}\centering \vspace*{4pt} \includegraphics[scale=0.06]{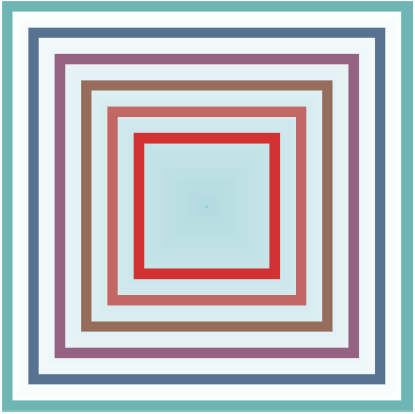}\\\small{PSCC 2020} \end{minipage} \hfill \small{Porto, Portugal --- June 29 -- July 3, 2020}}%
}

\begin{document}


\title{Differential Geometric Foundations for Power Flow Computations}

\let\bibsection\relax
\newcommand{\authors}[1]{#1}

\author{

\IEEEauthorblockN{Franz-Erich Wolter, Benjamin Berger, Alexander Vais}
\IEEEauthorblockA{
Leibniz Universität Hannover, Germany
}
}

\hyphenation{qua-dra-tic}
\hyphenation{span-ned}
\maketitle

\begin{abstract}
This paper aims to systematically and comprehensively initiate a foundation for using concepts from computational differential geometry as instruments for  power flow computing and research. At this point we focus our discussion on the static  case, with power flow equations given by quadratic functions defined on voltage space with values in power space; both spaces have real Euclidean coordinates. The central issue is a differential geometric analysis of the power flow solution space boundary (SSB, also in a simplifying way, called saddle node bifurcation set, SNB) both in voltage and in power space. We present different methods for computing tangent vectors, tangent planes and normals of the SSB and the normals' derivatives. Using the latter we compute normal and principal curvatures. All this is needed  for tracing  the orthogonal projection of points on curves in voltage or power space onto the SSB on points closest to the given points on the curve, thus obtaining estimates for their distance to the SSB. As another example how these concepts can be useful, we present  a new high precision continuation method for power flow solutions close to and on the SSB called local inversion of the power flow map from voltage to power space,  assuming  the  dimension of  power flow's Jacobean zero space, called KERNEL, is one. For inversion, we present two  different geometry-based splitting techniques  with  one of them using the aforementioned orthogonal tracing method. The other considers the power flow  map close to  the SSB  as a perfect  quadratic folding construction. Here the singular quadratic folding part is merely restricted  to one-dimensional kernel  spaces mapping the latter to euclidean rays. The  inversion  is then achieved by  an unfolding construction geometrically inverting the prior folding. Here accuracy of the unfolding  is benefiting from our splitting construction in a meta sense  restricting the singular part of the inversion essentially to a one-dimensional real square root operation. We sketch basic results on the local topology of the SSB and via topological analysis disprove the existence of fork type branching in planar sections of the SSB, that were numerically observed in a major report. Finally we indicate the relevance of geodesic coordinates for solutions set in power flow computing.

\end{abstract}
\begin{IEEEkeywords}
Power flow, Power system control, Computational Differential geometry, Local Inversion of Singular Maps
\end{IEEEkeywords}

\allowdisplaybreaks
\section{Introduction and Related Work}
Computations in the context of power flow and power
grid engineering have been to a vast extent essentially
applications of  tools from numerical analysis combined
 with various  types of classical engineering computations
 modified ad hoc for the equations under consideration.
 For those engineering problems, we have to analyze 
solution sets of non-linear equations, usually restricted by 
constraints often being nonlinear or causing additional 
non-linear  equations, and possibly varying with time.  
This paper will focus on the static case, but the concepts 
and results can be extended to the dynamic case to be
 presented in follow up papers. 

We are convinced of  the importance of understanding
 the geometric structure  of those solution sets for the 
following reasons: Theory from Riemannian and
 differential geometry helps  with understanding the
 local and the global structure  in a qualitative sense. 
 This insight, along with concepts from computational
 geometry, also yields results  of computations that are
 more precise and better organized. The solution sets 
relevant for engineers, typically defined  by 
constraints, are  natural  geometric objects, as they are
 implicitly defined Riemannian submanifolds  of 
Euclidean space  showing geometric structures 
inherited from  the surrounding space.
Over the recent decades, computational differential geometry
 has developed tools for subtle and precise computations
 on those submanifolds \cite{wolter2011computational}; those results also have engineering implications for the problem at hand,
which we intend to elaborate \cite{wolter2016MIT, wolter2017MIT, wolter2018NTU, wolter2018CGI, wolter2019NTU}. In power flow computing  there seem to be no systematic earlier  attempts to do so.

An AC power transmission system is usually modeled
as a graph. The vertices of the graph are called
buses and represent generators or consumers. The
edges are weighted with complex numbers and represent
transmission lines with associated admittances.
The state of an alternating current power transmission
system can approximately be captured by two sets of
variables indexed by the set of buses.   Complex variables
are usually split in their real and their imaginary part in
order to obtain pure real formulations presented in 
Euclidean  coordinates resulting in quadratic equations
giving several special properties which can be
exploited \cite{makarov2000properties}. This presentation being  extensively
used in research in power flow computing is very 
convenient for presenting our research and methods.
Therefore  we use it here. It is possible to  abstract
from the dynamical behavior of the system \cite{dobson1994irrelevance}.

This problem is presented  as  an analysis of   
the so-called static power flow equation, as described
in a seminal paper of Hiskens  \cite{hiskens2001exploring}. Even the
latter seemingly special  problem is  extremely 
complex and topic  of an awesome amount of 
ongoing research. 

Following  a presentation  in \cite{hiskens2001exploring}
we look at this problem in a notationally 
generalized   setting  as we want to keep the 
notations and descriptions simple  and short.  
Thus we have only
the \emph{voltage space} $V=\IR^n$, with vectors $v\in V$ and the
\emph{power space} $P=\IR^n$, with vectors $p\in P$ and the so called
\emph{power flow map} $F: V \funto P$ given by quadratic
forms. All the engineering relevant requirements
 may  then be  obtained  by considering  particular
coordinates  for vectors in $V$ and $P$-space,  say  referring  to
 features such as  active load and power
injection, active and reactive coordinates in power space.
Of course there are also additional
nonlinear constraints such as  those for voltage 
magnitudes and additional  ones  that may {e.g$.$}
result from investigating special situations. 
Using  the typically  employed  complex notation
or employing various different  types of variables
would confuse the explanation of our methods 
as those are  independent of the special interpretation of the coordinates involved.
 
In power flow applications, the function $F$ is not just any quadratic function, but has a symmetry that allows to declare one bus as the ``slack bus'', that is to set one voltage phasor to
any nonzero value, typically $1$, without loss of generality. This eliminates two variables, but we lose the property that the functions under consideration are homogeneous. We chose to not make use of this symmetry and to formulate our equations with respect to arbitrary homogeneous polynomials.

For explaining  our  concepts we focus on  some  major  fundamental aspects of problems causing basic  difficulties  in  many sample problems in  power flow computing. Once the formulas are available they may  be applied to particular equations stemming from power flow  computing. We did this  for sample  problems involving  concrete sample  networks  with up to 220 dimensions, taken from \cite{tcarchive}.

In some  technical  and partly  differential geometric aspects works in power flow computing described  in  \cite{alvarado1994computation, dobson1993computing, dobson1993new, dobson2003distance}  are close to ours.  Albeit  restricted  to  the computationally more  accessible power space, these  works use and benefit  from differential geometric concepts for computing  nearest points  on and  distance estimations to saddle point bifurcations.  There are major differences in our computations:  We present  distance and nearest point computations  including the  harder case  of distances in  voltage space.  We  also include  (orthogonal) tracing  of points  nearest on the singular set  to points  moving on a space curve.

Recent works such  as  \cite{yao2018improving} on finding feasible solution sets points of maximal and minimal distance  to SSB indicates current  interest in distance computations of operation points to  the SSB. \cite{yao2018improving} searching  in  feasible  solution sets points  of  maximal or minimal distance to SNB of SSB proves the need for distance computations of operation points to the SSB. Oppose to us, \cite{yao2018improving} seems not to use of maximal principal curvature--or equivalent second-order surface properties of the SSB, needed for local distance minimality (cf. fig. \ref{figKruku}). This would require observing  geometric second-order input from both SNB and the feasible manifold.

Although the question of systematically solving the power flow equation close to the singular boundary is considered important, we did not see works employing an approach similar to our local inversion technique, which separates the difficult-to-invert part into a one-dimensional subspace.

Discussing applications of our new methods on special EE sample problems is so far still in the beginning and have not been our focus, as we  are still developing and improving basic tools. We also  feel that  the more global vision of our concepts outlined with some details and many figures in our work \cite{wolter2019differential} is more important and with details and proofs should have a broader long time impact. In \cite{wolter2019differential} we prove, discuss and illustrate basic results of the local topological SSB structure, being here only briefly sketched. Likewise here we only descriptively indicate the relevance of geodesic coordinates for constraint solution sets 
and refer for details to \cite{wolter2019differential}.


\section{Geometric Entities on the SSB}
\label{secSSB}
The set of points where the Jacobian of the power flow map is singular is called the \textit{\textbf{solution space boundary}} (SSB) or bifurcation set \cite{hiskens2001exploring}. Its physical relevance is that the solution of the power flow equation $F(v)=p$ cannot be continued continuously beyond the SSB, leading to instability of the system. Understanding the SSB is important in order to keep a safe distance from it.

The SSB \emph{normal in power space}, $N_P$, at a point is simply the left eigenvector of the differential for the eigenvalue $0$:
\begin{align}
(\D F)\transp N_P &= 0 \\
N_P\transp N_P&=1.
\end{align}
The columns of $(\D F)$ are all orthogonal to the normal, so they span the tangent space.

The vector $N_P$ is in the zero space of $(\D F)\transp$,  which we  call \emph{kernel}. We mostly have the generic  case where the dimension of this zero space is $1$,  allowing us to identify it with a non-oriented unit vector, called kernel vector. 
Because $N_P$ is in the kernel of $(\D F)\transp$, we will also call it ${\tilde k}$, whereas $k$ will be the kernel vector of $\D F$, thus being much harder to compute than $N_P$.

In voltage space, things are more complicated. Because the SSB is an iso-surface where $\lambda_0$, the smallest eigenvalue of $(\D F)$, is zero, the \emph{unit normal in voltage space} $N_V$ is collinear with the gradient of $\lambda_0$.

Differentiating the equations expressing that $(\lambda_0, {\tilde k})$ is an eigenpair
with respect to a voltage variable $v_i$ at $\lambda=0$, we obtain a linear equation system with unknowns $\dquot{\lambda_0}{v_i}$ and $\dquot{{\tilde k}}{v_i}$:
\begin{align}
(\D F)\transp\dquot{{\tilde k}}{v_i} - {\tilde k} \dquot{\lambda_0}{v_i}&= -\dquot{(\D F)\transp }{v_i} {\tilde k}  \label{vNormal}\\
\dquot{{\tilde k}\transp}{v_i} {\tilde k}&= 0. \nonumber
\end{align}
Thus we obtain both the \emph{(unnormalized) normal vector in voltage space} and the \emph{derivatives of the normal vector in power space} (provided the rank of $\D F$ is $n-1$ and hence the zero eigenvalue is simple). We are also interested in curvatures of the SSB in voltage space. So in order to get the derivatives of the unnormalized normal in voltage space, we need to differentiate again with respect to a voltage direction $v_j$:
\begin{align}
\label{evsecder}
(\D F)\transp\ddquot{{\tilde k}}{\der v_i\der v_j} - & {\tilde k} \ddquot{\lambda_0}{\der v_i\der v_j}
= -\ddquot{(\D F)\transp }{\der v_i\der v_j} {\tilde k} -\dquot{(\D F)\transp }{v_i} \dquot{{\tilde k}}{v_j}\nonumber \\
&- {\left(\dquot{\D F - \lambda_0 \ONE}{v_j}\right)}^\top\dquot{{\tilde k}}{v_i}+ \dquot{{\tilde k}}{v_j} \dquot{\lambda_0}{v_i}\\
&\ddquot{{\tilde k}\transp}{\der v_i\der v_j} + \dquot{{\tilde k}\transp}{v_i} \dquot{{\tilde k}\transp}{v_j} = 0. \nonumber
\end{align}
By inserting the previously calculated values of $\dquot{\lambda_0}{v_i}$ and $\dquot{{\tilde k}}{v_i}$, we get a linear equation system for
the second derivatives $\ddquot{\lambda_0}{\der v_i\der v_j}$ and $\ddquot{{\tilde k}}{\der v_i\der v_j}$. Since this equation system has the same matrix as \eqref{vNormal} (and the same as the matrices needed for higher derivatives), regardless of $i$ and $j$, it pays off to invert the matrix once. Once the matrix has been inverted, derivatives of the same order can be computed in parallel. 

This is a systematic approach to compute arbitrary derivatives of the normal vectors. However, in our actual numerical experiments where we only needed the normals and their first derivatives, we used ideas presented in the following which allowed us to obtain the first derivatives of the normals without using derivatives of eigenvalues and eigenvectors.
In our implementation, we made no use of derivatives of eigenvalues of $\D F(p)$, not even for computations  of tangent vectors and the normal on the SSB in voltage space. Instead of that we were  using  \eqref{Lij}  extended with the  additional  equation $\dot k\transp k=0$ to compute tangent vectors and from these the normal, see \cite{gruhl}.

First derivatives of the normal are needed to define \emph{normal curvatures} in various tangential directions $\dot c$. Let $N$ be the unit normal vector in either power or voltage space. 
The normal curvature $\kappa_N(\dot c)$ in the unit direction $\dot c$ is defined as 
\begin{align}
\kappa_N(\dot c) &= W(\dot c)\cdot \dot c, \label{wg}
\end{align}
where $W$ is the negative differential of the unit normal, $W=- \D N$, called \emph{Weingarten map} or \emph{shape operator}. 

\skipall{
In terms of the above directional derivatives, which only give us non-unit normals $\tilde N$, the normal curvature is
\begin{align}
\kappa_{\frac{\tilde N}{\abs {\tilde N}}}(\dot c)&= \left(\frac{\tilde N \left( \tilde N \cdot \dquot{\tilde N}{\dot c} \right)}{\abs{\tilde N}^3} - \frac{\dquot{\tilde N}{\dot c}}{\abs{\tilde N}}\right) \cdot \dot c.
\end{align}
}
If $\dot c$ is actually the tangent vector of an arc-length parametrized surface curve $c$, then using  Meunier’s theorem there is a simpler way to compute the normal curvature:
\begin{align}
\kappa_N(\dot c)&= N \cdot {\ddot c}. \label{nks}
\end{align}
Only the component of ${\ddot c}$ perpendicular to the surface is relevant in \eqref{nks}, and this normal component only depends on $\dot c$ and no other details of the curve.

We can use this for computing the shape operator without the need to differentiate the normal vector via the subsequent way, inspired by \cite{Wolter1992c}: 
Choose $\frac{n^2-n}{2}$ suitable unit tangent directions $\dot c_i$. 
Since the curves $c_i$, assumed to be parametrized proportional to arc length, run inside the surface, the normal component of each $\ddot c_i$ is uniquely determined by $\dot c$ (see \eqref{ddc}) and can be used in \eqref{nks} to compute the normal curvatures in the directions $\dot c_i$. By inserting all the known normal curvatures into $\frac{n^2-n}{2}$ instances of \eqref{wg}, one instance for each $i$, we get a linear equation system for the components of the matrix representation of $W$:
\begin{align}
W(\dot c_i)\cdot \dot c_i = \kappa_N(\dot c_i).
\end{align}

This is an equation system with $\frac{n^2-n}{2}$ equations and the same number of unknowns; $W$ is self-adjoint with respect to the \emph{Riemannian metric tensor} or \emph{first fundamental form} $g$, so only $\frac{n^2-n}{2}$ entries are needed to determine it. The metric tensor is used to define the inner products and thereby lengths and angles of tangential vectors. In our case, the manifold is embedded in a Euclidean space and so $g$ is simply the restriction of the ambient Euclidean metric to the tangent space: $g(\dot c_i, \dot c_j)$ is simply $g_{ij} := \dot c_i \cdot \dot c_j$.
In particular $W_{ik} = -\sum_j g^{ij}L_{jk}$ where the $g^{ij}$ are the components of the inverse of the matrix $(g_{ij})$. The $L_{jk}$ are the components of the second fundamental form, which form a symmetric matrix.

By cleverly choosing the directions $\dot c_i$ , we can ensure that the equation system is sparse and efficiently solvable: Choose the first $n-1$ directions as the standard basis of the tangent space, and the remaining directions as the sums of each pair of distinct basis vectors.
The equation system for the components of $L$ then consists of the equations:
\begin{alignat}{3} \label{Lij}
L_{ii} &= \kappa_N\left(\dot c_i\right) \quad &&\text{for } 1\leq i<n \\
L_{ii} + 2 L_{ij} + L_{jj} &= \kappa_N\left(\dot c_i + \dot c_j\right) \quad &&\text{for } 1\leq i< j<n.\nonumber
\end{alignat}
Since $\kappa_N$ is applied to non-unit vectors here, we must define it to behave as a quadratic form: $\kappa_N(s v) = s^2\kappa_N(v)$.

It remains to find a possible second derivative $\ddot c$ (we're only interested in the normal part) of an arc-length parameterized curve $c$ going in the direction $\dot c$, let us first consider some algebraic properties of the Jacobian of a quadratic function $F:~\IR^n~\funto~\IR^n$.
Such a function can always be written using symmetric matrices $A_i$, $1\leq i\leq n$:
\begin{align*}
	F(v) =  \begin{pmatrix}
		F_1(v)\\
		\vdots\\
		F_n(v)
	\end{pmatrix} = \begin{pmatrix}
	v^\top A_1 v\\
	\vdots\\
	v^\top A_n v
	\end{pmatrix},
\end{align*}
Because these matrices are symmetric, the Jacobian is simply:
\begin{align*}
\D F(v) = 2 \cdot
\begin{pmatrix}
	v^\top A_1\\
	\vdots\\
	v^\top A_n
\end{pmatrix}
\end{align*}
and one easily checks that
\begin{align}
\label{eq27}
(\D F(x)) v &= 
\skipall{2 \cdot
\begin{pmatrix}
	x^\top A_1 v\\
	\vdots\\
	x^\top A_n v
\end{pmatrix}= 2 \cdot
\begin{pmatrix}
	v^\top A_1 x\\
	\vdots\\
	v^\top A_n x	
\end{pmatrix} \nonumber\\
& = }(\D F(v)) x.
\end{align}
The Hessian of a quadratic form is constant:
\begin{align*}
\Hs F(x) = \dquot{\D F(x)}{x} = 2 \cdot
\begin{pmatrix}
	A_1\\
	\vdots\\
	A_n
\end{pmatrix}.
\end{align*}
This implies the following:
\begin{align}
v\transp\left(\Hs F(x)\right) =& (\D F)(v). \label{eq28}
\end{align}

\skipall{The function $F$ is a homogeneous quadratic form, hence the SSB is composed of cones: For any $c$ so that $\D F(c)$ is singular, the vector from the origin to $c$ is tangential to the SSB.}

With this, we can obtain the \emph{second derivatives of the curves $c$ in voltage space} by solving the subsequently derived linear systems \eqref{Lija}, \eqref{Lijb}.
For this with $k$ being the kernel vector of $\D F$, we differentiate $(\D F (c)) k = 0$ twice, applying \eqref{eq28} and \eqref{eq27}:
\begin{align} 
(\D F (k)) \dot c + (\D F (c)) \dot k &= 0.\label{Lija} \\
(\D F (k)) \ddot c  + (\D F (c)) \ddot k &= -2 (\D F (\dot c) ) \dot k.\label{Lijb}
\end{align}
This is an under-determined equation system for the unknowns $\ddot c$ and $\ddot k$. Hence we make it uniquely solvable by requiring the kernel to be normalized (differentiating $k \cdot k = 0$ twice gives $\ddot k \cdot k = - \dot k \cdot \dot k$) and assuming, since we're only interested in the normal part of $\ddot c$, that $\ddot c = \kappa N_V$ for some $\kappa\in\IR$.
Finally, we arrive at an equation system with $n+1$ unknowns and the same number of equations:
\begin{align}
\label{ddc}
\begin{pmatrix}
(\D F (k)) N_V & \D F (c) \\
0 & k\transp
\end{pmatrix}
\begin{pmatrix}
\kappa \\ \ddot k
\end{pmatrix}
=
\begin{pmatrix}
-2 (\D F (\dot c) ) \dot k \\ - \dot k \cdot \dot k
\end{pmatrix}.
\end{align}
To determine $\dot k$, we add the condition that $k$ remain normalized, expressed as $k\cdot \dot k=0$, to \cref{Lij}:
\begin{align}
\begin{pmatrix}
\D F (c) \\
k\transp
\end{pmatrix}
\dot k=
\begin{pmatrix}
- (\D F(k))\dot c \\ 0
\end{pmatrix}.
\end{align}
These are $n+1$ equations for $n$ unknowns, but they always admit a solution since the first $n$ equations are linearly dependent.

In our specific setting, this completes the computation of normal curvatures for curves on the SSB, given their tangent vectors.  
In  the  preceding computations and considerations   related to  computing $W$ or equivalently $\D N_V$  (starting with \eqref{wg}),  we were using the assumption that normal curvatures for tangent directions are known. Combining these 
results then we can now concretely compute $\D N_V$ and $W$ for the SSB  in voltage space.  We used  this  method  in our implementation for all numerical experiments involving $\D N_V$ and $W$.

The \emph{second-derivatives of the images of curves on the image of the SSB in voltage space} require slightly more effort. For this, we use the fact that the first derivative of the image $F(c)$ with respect to the curve parameter $t$ is $(\D F(c)) \dot c$. If we differentiate this again and then apply \eqref{eq28}, we have
\begin{alignat*}{3}
 \ddquot{F(c)}{\der t^2} &=\dquot{(\D F(c)) \dot c}{t} 
&&=(\D F(\dot c))\dot c + (\D F(c)) \ddot c.
\end{alignat*}

When $n$ is large and we are interested in only a few normal curvatures in power space, say one in direction $(\D F(c)) \dot c$, it is more appropriate to not compute the complete shape operator using \eqref{Lij}, but rather use the following basic classical formula for applying the shape operator, given in the local coordinates provided by interpreting $F$ as parameterization of power space by voltage space:
\begin{align} \label{eq15}
W(\dot c)_i = g^{-1} \tilde L \dot c,
\end{align}
restricted to the tangent space of the SSB image in power space. 
Here $g = (\D F)\transp (\D F)$ is the first fundamental form related to $F$, which is given by a sparse symmetric matrix. The matrix $\tilde L$ yields the second fundamental form of the SSB-image when restricted to tangential space. 
Its components are $\tilde L_{ij}=\ddquot{F}{\der v_i \der v_j} \cdot N_P$, which is also a sparse symmetric matrix.
The matrix $g$ is singular on the SSB. But we can avoid having to invert the singular $g$ by multiplying it to the left side:
\begin{align}
g (W(\dot c)) = \tilde L \dot c. \label{sswm}
\end{align}
To make this linear equation system for $W(\dot c)$ more palatable to numerical solvers, we may want to ensure the right hand side does not have a component in the unit direction $k$ of the kernel of $g$ by writing
\begin{align}
(g + \epsilon k k\transp) (W(\dot c)) = (\ONE - k k\transp) \tilde L \dot c.
\end{align}
for some $\epsilon>0$.
The solution $W(\dot c)$ may not be tangential in voltage space, but since we are only interested in its image $(D F)(W(\dot c))$ in power space, this does not matter.

Since in our context $g$ originates from the adjacency matrix $\D F$ of a mostly planar graph with node degree in practice bounded by some small $k\in\IN$, we suggest solving this equation system using an algebraic multigrid technique. This should take $O(nk^2)$ time.

We will also need \emph{principal curvatures} for distance estimation purposes. These are the eigenvalues of the Weingarten map.
Since matrices of the first fundamental form $g$ and of the second fundamental form $L$ are known, we can calculate some or all of these by solving the sparse generalized eigenvalue problem
\begin{align}
g v = \kappa_N(v) L v.
\end{align}
This yields the corresponding principal curvature direction $v$.

\section{Local Inversion}
\label{secLocInv}

Given a curve $p$ in power space, and a solution $p(0)=F(v(0))$ of the power flow equations for its starting point $p(0)$, one would often like to know the particular connected component $v$ of the preimage of $p$ in voltage space that contains $v(0)$. Standard continuation methods become unstable near the SSB, where the Jacobian is nearly singular, because conceptually, they require multiplication of the inverse Jacobian by $p^\prime(t)$ to obtain $v^\prime(t)$. 

To solve this problem, we have devised two alternative approaches \cite{hein}. Both are based on the idea of splitting the representation of the point $v(t)$ into a pair consisting of a point $q(t)$ on the SSB and a distance $d(t)$ from $q(t)$ in a particular direction $w(q(t))$, so that $v = q + d w(q)$. Depending on the approach, the direction $w(q)$ is chosen to be either the kernel of the differential at $q$, or the surface normal $N_V(q)$ (See \eqref{figLI} for a sketch). We do not elaborate on how to obtain the initial values of $w$ and $q$; for $w=N_V$ one can use orthogonal projection (\eqref{secOrtho}), whereas for $w=k$ we have so far used Newton's method to obtain a solution, or gradient methods in case  some  point $q^*$ on SSB close to $v$ was known. In this case we used a gradient  descent method to trace a curve on the SSB starting at $q^*$ that would end up in a  location  $q$ where the kernel would be collinear with the segment joining $q$ and $v$.  

Both approaches allow us to stably compute $v^\prime(t)$ from $p^\prime(t)$ even in the vicinity of the SSB. Because the terms of $F$ are at most quadratic, its Taylor series contains at most three terms and can be used to express the dependency of $v^\prime(t)$ on $p^\prime(t)$ by a well-conditioned linear equation system.

Both approaches should be combined: The kernel is easier to compute, needing fewer derivatives and usually providing more accuracy, but may become tangential to the SSB. If it becomes tangential, it is unreliable for the task of representing $p(t)$ and the algorithm should switch to using the normal for $w(q)$, by finding an orthogonal projection of $p(t)$ onto the SSB.

To derive the continuation method, we differentiate the equation $p = F ( q + d \cdot w (q) )$
with respect to $t$ (denoted by a dot over the variable):
\begin{align}
\dot p = (\D F ( q + d \cdot w ) ) \left(\dot q + \dot d \cdot w + d \cdot \dot w \dot q \right).
\end{align}
We want to use that $F$ is quadratic. (In the following, the Jacobian $\D F$ and the Hessian $\Hs F$ are always implicitly evaluated at the location $q$ unless an explicit argument is specified.) Then the second-order Taylor approximation is actually exact:
\begin{figure}[tp]
\fbox{
	\parbox{7.5cm}{
		\center
		\includegraphics[width=7cm]{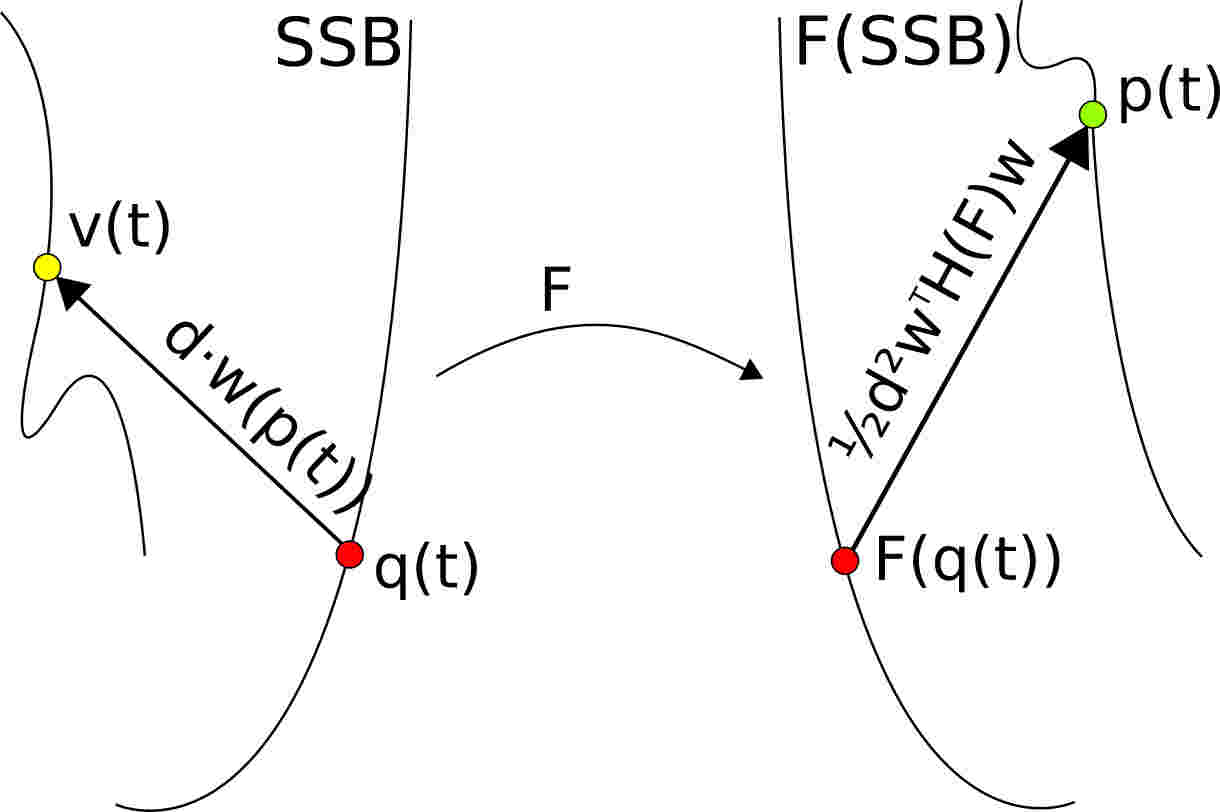}
		\caption{Calculating the local inverse of a curve using the split representation. H denotes the Hessian.}
		\label{figLI} 
	}
}
\end{figure}
\begin{align}
\label{liw}
p = F(q) + d \cdot (\D F) w + \frac{d^2}{2} \cdot  w \transp (\Hs F)  w ). 
\end{align}
The linear term in \eqref{liw} will vanish if we use the kernel $k(q)$ (or short $k$) of the differential at $q$ as our choice for $w(q)$, leading to the simplified equation
\begin{align}
\label{lik}
p = F(q) + \frac{d^2}{2} \cdot k \transp (\Hs F) k.
\end{align} 
This case is illustrated in \eqref{figLI}. We also provide an animation under \url{https://www.dropbox.com/s/nlu2bp1nywmkb8c/folding.mp4} that illustrates the idea of how the power flow map acts locally like a quadratic fold at the SSB, in the generic case.
Differentiating \cref{lik} with respect to $t$ yields
\begin{align}
\label{likd}
\dot p = (\D F ) \dot q + d \dot d \cdot k \transp (\Hs F) k + d^2 \cdot k \transp (\Hs F) \dot k.
\end{align} 
We used that the derivatives of the Hessian vanish. 

We can turn \eqref{likd} into a linear equation system for $\dot q$, $\dot k$ and $\dot d$ by adding the condition that $\dot q$ be tangential and $k$ remain the kernel:
\begin{align}
\begin{pmatrix}
\D F & d^2 \cdot k \transp (\Hs F) & d \cdot k \transp (\Hs F) k \\
k\transp (\Hs F) & \D F & 0 \\
0 & k\transp & 0
\end{pmatrix}
\begin{pmatrix}
\dot q \\ \dot k \\ \dot d
\end{pmatrix}
=
\begin{pmatrix}
\dot p \\ 0 \\ 0
\end{pmatrix}. 
\end{align}
This still becomes ill-conditioned near the SSB where $d$ is small, but the problem is confined to the unknown $\dot d$, and we know that the length of the vector $p-F(q)$ scales exactly quadratic with $d$ (\eqref{lik}), providing an alternative means to update $d$. This exploitation of the quadratic folding behavior of $F$ 
 presented in the geometric splitting construction restricting the singular part of the map to the family of 1-d kernel vector lines  each  mapped to quadratically  scaled  Euclidean rays, lead   here to   geometric natural  unfolding construction providing  a high precision inversion. 

According to \cref{eq28}, $k\transp (\Hs F(q)) = \D F(k)$ for all $q$. This simplifies the formulas used here in several places.

We use this equation system to implement a continuation method that traces how the preimage $v = q + d \cdot k$ of $p$ evolves as $p$ changes in the direction $\dot p$. There are some problems with this that need to be addressed: First, as the curve $q$ is constructed, it may deviate from the SSB due to numerical inaccuracies. If that happens, it should be corrected by projecting it back onto the SSB along the direction $w(q)$. Second, we found in experiments that the simplification employed in \cref{lik}, while valid when using exact arithmetic, leads to numerical errors that can be dramatically reduced in exchange for the rather small effort of using the full \cref{liw}. Third, it may not always be possible or reliable to represent $v$ as $q + d \cdot k(q)$, namely if the kernel is (nearly) tangential. In that case, it would be more appropriate to use the normal instead of the kernel for $w$. A suitable point $q$ so that $v$ can be expressed as $q + d \cdot N_V(q)$ can be found using orthogonal projection (see \eqref{secOrtho}). Using the normal should be avoided when the kernel is available for representing $v$ because the differential of the normal is much more expensive to compute.

For these reasons, we should also derive the equation system for $\dot q$ and $\dot d$ when the full \eqref{liw} is used. Differentiating \cref{liw} with respect to $t$, we get:
\begin{align}
\label{liwd}
\dot p =& (\D F) \dot q 
+ \dot d \cdot (\D F) w + d \cdot \dot q \transp (\Hs F) w \\
& + d \cdot (\D F) \dot w 
+ d \cdot w \transp (\Hs F) k + d^2 \cdot w \transp (\Hs F) \dot w \nonumber.
\end{align} 

To handle the case $w \eqdef k$, we can as above derive a linear equation system from this, solving which tells us $\dot q$, $\dot d$ and $\dot k$, given $\dot p$. For the case $w := N_V$   we  use the directional derivative of  $N_V (q)$ in the direction $\dot q$, which is given by $(\D N_V (q))\dot q$.  For this, we have so far used and tested our methods in \ref{secSSB} for  explicitly computing the Weingarten matrix. (We now consider an improved geometric method with much lower complexity, which still needs to be tested.)

\subsection{Numerical Results}
We tested the local inversion algorithm on power network configurations from the power system test case archive of the University of Washington \cite{tcarchive}. In particular, we compared the precision of three of our methods: The method that uses the kernel and omits the linear term of the Taylor expansion that is theoretically zero, the method that uses the kernel but includes the linear term, and the method using the normal vector. We start with curves in voltage space, map them through $F$ into power space and invert the results back into voltage space using the three algorithms described above. By comparing the result with the original curve, we can estimate the precision, which is shown in \eqref{tabPrecision} (taken from \cite{hein}).

We  point out that all these test implementations  as  well  as  all  other ones  were  very far away from  being  optimized. They were rather  first  experimental  proofs of concept that  all  those computations are possible. 

\begin{table*}[hbt]
	\centering
	Step size $10^{-9}$ \\
	\begin{tabular}{|c|c|c|c|}
		\hline 
		\textbf{\# of buses} & \textbf{Kernel w/o lin.} & \textbf{Kernel with lin.} & \textbf{Normal}\\ 
		\hline 
		14 & $2.38774883365\cdot 10^{-5}$ & $2.37706864821\cdot 10^{-5}$ & $2.14982954847\cdot 10^{-5}$\\ 
		\hline 
		30 & $7.05835279589\cdot 10^{-9}$ & $7.05834973039\cdot 10^{-9}$ & $7.97487174331\cdot 10^{-7}$ \\ 
		\hline 
		57 & $1.64770875908\cdot 10^{-7}$ & $1.64770753583\cdot 10^{-7}$ &$3.46424134738\cdot 10^{-6}$ \\ 
		\hline 
		118 & $6.71182537764\cdot 10^{-8}$ & $1.07999417455\cdot 10^{-13}$ & $1.9659196324\cdot 10^{-7}$\\
		\hline 
	\end{tabular} \\
	Step size $10^{-7}$ \\
	\begin{tabular}{|c|c|c|c|}
		\hline 
		\textbf{\# of buses} & \textbf{Kernel w/o linear} & \textbf{Kernel with linear} & \textbf{Normal}\\ 
		\hline 
		14 & $2.41940697575\cdot 10^{-5}$ & $2.38882856096\cdot 10^{-5}$ & $1.20966340549 \cdot 10^{-4}$\\ 
		\hline 
		30 & $7.05835852083 \cdot 10^{-6}$ & $7.05835718598\cdot 10^{-6}$ & $7.97486759134 \cdot 10^{-5}$ \\ 
		\hline 
		57 & $1.64760675278 \cdot 10^{-4}$ & $1.6477270041\cdot 10^{-5}$ & $3.34938440007 \cdot 10^{-4}$ \\ 
		\hline 
		118 & $5.37431102992 \cdot 10^{-7}$ & $1.1515965505\cdot 10^{-10}$ & $2.0439378252 \cdot 10^{-4}$\\
		\hline 
	\end{tabular}
	\caption{Average distance between original curve and inverted curve after 100 steps with step size $10^{-9}$ (upper table) and $10^{-7}$ (lower table) }
	\label{tabPrecision}
\end{table*}

Using the linear term in the kernel method pays off in precision, for almost no additional runtime cost.

A comparison of execution times for a single step is displayed in \eqref{tabRuntime}. 
\begin{table}[hbt]	
	\centering
	\begin{tabular}{|c|c|c|}
		\hline 
		\textbf{\# of buses} & \textbf{Kernel} & \textbf{Normal}\\ 
		\hline 
		14 & 0.278 & 0.324\\ 
		\hline 
		30 & 3.329 & 6.492 \\ 
		\hline 
		57 & 42.047 & 81.992\\ 
		\hline 
		118 & 225.323 & 439.380\\
		\hline 
	\end{tabular}
	\caption{Runtime for a single step of the algorithm in seconds on a Laptop with an Intel i7 7700HQ processor }
	\label{tabRuntime}
\end{table}

\section{Jacobean   Hessian  of  Local  Inverse}
\label{secJaco}

In \ref{secLocInv} we   presented   well-tested methods tracing  for given curves in power space close to and on the SSB  local inverse  (pre-image)  curves  in voltage space by  computing its tangent vectors. This gave us all first-order directional derivatives of  the local  inverse $F^{-1}$ in the respective local set in power space, thus especially  all partial derivatives  hence the Jacobean of that  local inverse $F^{-1}$.

In \cite{wolter2019differential} formula (67) we presented a method (that we have tested) for directly computing the second-order derivative $\ddot v(t)$ of a preimage curve $v(t)=F^{-1}(p(t))$. One could also compute second-order derivatives of preimage curves $v(t)$ of given curves 
\begin{equation} \label{eqK}
p(t)=F(v(t))
\end{equation}
using differentiating \eqref{eqK} twice with respect to $t$ yields
\begin{equation*}
\ddot p(t) = \dot v^T H(H(v(t))) \dot v(t) + DF(v(t)) \ddot v(t) 
\end{equation*}
thus
\begin{equation} \label{eq26}
DF(v(t)) \ddot v(t)  = \dot v^T H(v(t)) \dot v(t) - \ddot p(t) 
\end{equation}
We avoided this equation \eqref{eq26} in case $D(F(v(t)))$ is ill conditioned. But having a decent inverse $D(F(v(t)))^{-1}$ one could use this equation
without the need of computing  and using  the  full  mostly non-sparse  inverse of the Jacobean.   
In our case with the Jacobean $D(F(v(t)))^{-1}$ available from
\ref{secLocInv} we yet preferred the way described in \cite{wolter2019differential} formula (67) as to obtain $\ddot v$ from $\ddot p$ as we often only need second derivatives of selected image curves.

Since otherwise in this situation with
$DF$  in  \eqref{eq26} being ill conditioned we 
might alternatively  multiply  equation  \eqref{eq26} from the left with  $DF^{-1}$ as to  get an equation   for $\ddot v$.    Even if
that would work we would need 
computing  the full  non sparse inverse  $DF^{-1}$ which we wanted to avoid. Hence we preferred the   direct method in \cite{wolter2019differential} formula (67) for second order derivatives which also seems to be more accurate as well. 
\footnote{Clearly one also could approximate less  accurately   the second derivative by  difference methods	from the first derivatives.}

We proceed to compute the Hessian $HF^{-1}(p_0)$.
For computing the Hessian $HF^{-1}(p_0)$ we already have the diagonal second-order partials $\partial_{p_i}^2 F^{-1}(p_0)$ 
as we know how to compute second-order  derivatives of  pre-image curves and need the mixed partials $\partial_{p_i}\partial_{p_j}F^{-1}(v_0)$.
 For this, we define a two-variable embedding map $\varphi(s,t) = (s p_i + t p_j)$ in power space with basis vectors $p_i, p_j$, next we define $\psi(s,t) := F^{-1}(\varphi(s,t))$ and three curves $v_1(s):= \psi(s,0)$, $v_2(t) := \psi(0,t)$ and $v_3(r) := \psi(r,r)$.
The second-order derivatives of these three curves which we can compute yield the $(2,2)$ Hessian $H\psi(s,t)$ with four vectorial elements, where two diagonal ones are the second derivatives of $v_1,v_2$.
Knowing those the second derivative of $v_3(r)$ yields the non-diagonal element using here that the second derivative of $v_3(r)$ fulfills the quadratic form $(s,t) H\psi (s,t)^T$ with $s=t=1$ thus giving access to the
$\partial_t \partial_s \psi$ providing the wanted mixed partial $\partial_{p_i}\partial_{p_j} F^{-1}(p_0)$.

\section{Orthogonal Projection}
\label{secOrtho}

\begin{figure}[tp]
\fbox{
	\parbox{7.5cm}{
		\center
		\includegraphics[width=7cm]{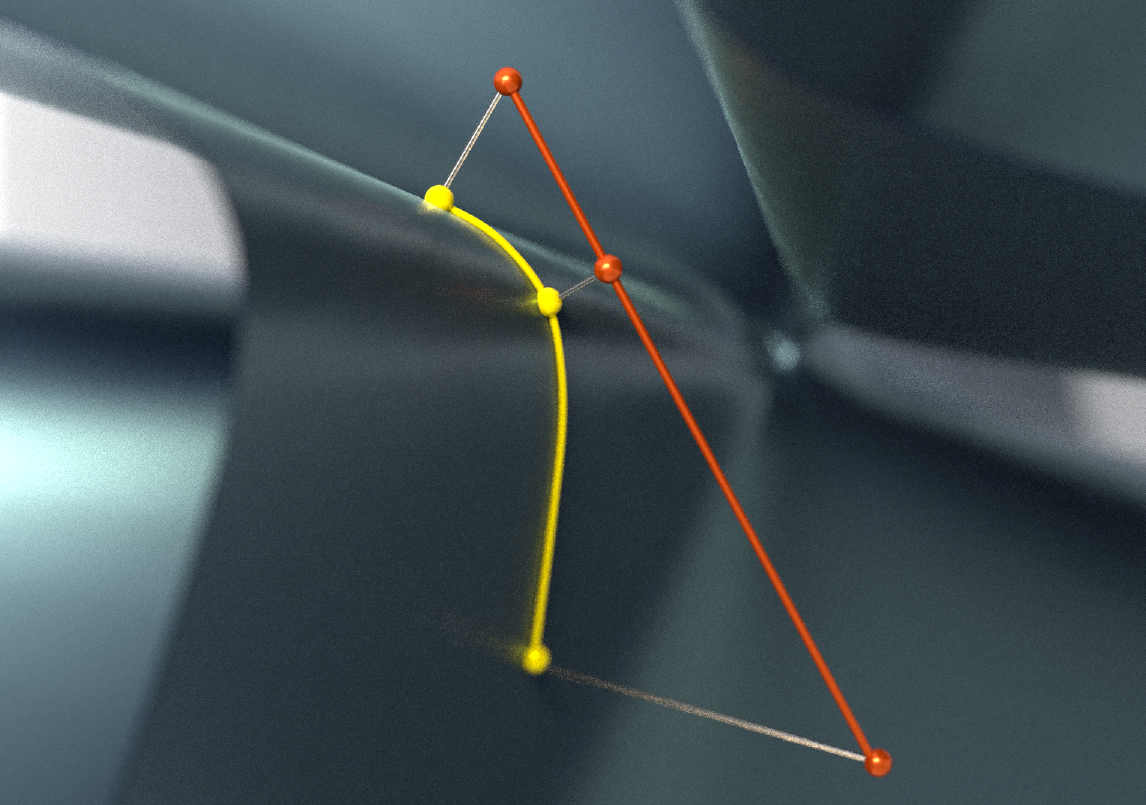}
		\caption{Orthogonal projection of a curve onto an SSB.}
		\label{figOproCurve} 
	}
}
\end{figure}

For some purposes, such as our local inversion technique when using the normal vector, it is required that, given a point $v$, we find a point $q$ on the SSB so that the normal at $q$ points along the direction $v-q$. Since the SSB is the implicit surface where the absolutely smallest eigenvalue $\lambda$ of the power flow differential $\D F$ is $0$, we can use the approach described in the following to find that orthogonal projection $q$. Also of interest are orthogonal projections of entire curves onto the SSB \cite{pegna1996surface} (see \eqref{figOproCurve}).

For this, we can use the subsequent   fairly  delicate method searching for a locally minimal distance projection  $q$ of  
space point $v$.  We also successfully tested simpler gradient descent methods 
tracing curves on the SSB   starting in a point  $q_*$ near to $v$ that would end up in a point  $q$,  such that $(q-v)$  would be collinear with the normal $N_q$ on the SSB,  with SSB either in voltage  space  or in 
power space, the latter  being the easier case.

\subsection{An Algorithm for Computing Orthogonal Projections of Points onto the SSB}
\begin{figure}[tp]
\fbox{
	\parbox{7.5cm}{
		\center
		\includegraphics[width=5cm]{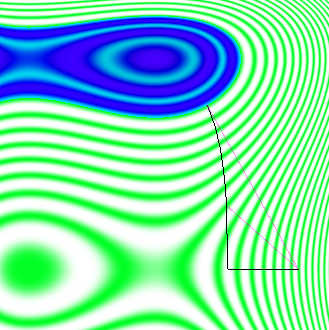}
		\caption{2D-illustration of the ODE for finding orthogonal projections onto a line (that is not SSB-like). The isolines have been visualized.
		The normal to the isoline at $t=0$ as well as a normal to an isoline halfway between $t=s$ and $t=0$ are marked.
		The black line is the curve $r$.}
		\label{figOpro2d} 
	}
}
\end{figure}
Let $\lambda:\IR^n\funto \IR$ be a function whose zero set is the surface that we want to project onto, such as the SSB with $\lambda$ being the smallest eigenvalue of $\D F$, and let $v$ be the point that we want to project.
First, we evaluate $s = \lambda(v)$. Then, we compute a parameterized curve $r$ so that $r(s) = v$, $r(0)=q$ and for all $t$ between $0$ and $s$, the normal to the iso-surface with $\lambda(r(t))=t$ and the direction $r(t) - v$ are linearly dependent. See \cref{figOpro2d} for an illustration of the idea.
The condition for lying on the isosurface is equivalent to the existence of a kernel $k$ of $\D F - \lambda \ONE$. 
We make the generic assumption that the gradient of $\lambda$ does not vanish\footnote{This is a rather safe assumption if the staring point is near the SSB and we have the generic case that the SSB is locally a manifold. Also, we conjecture that, apart from the initialization, our algorithm automatically avoids to approach such degenerate situations, should they occur.}, so that $r(t)-v$ can always be written as a multiple of it:
\begin{align} 
d(t) \cdot (\D \lambda(r(t)))\transp & = v - r(t) .
\end{align}
The expressions $\D \lambda$ (and its directional derivatives needed below) may be calculated analogously to \eqref{vNormal}.
Then, we trace the curve $r$ from $t=s$ to $t=0$ while maintaining this relation to stay true. For this, we differentiate with respect to $t$ (denoted by a dot on top of symbols) and add the condition $\dot \lambda=(\D \lambda(r))\dot r = \sign{s}$:
\begin{align} \label{orthoODE}
\begin{pmatrix}
d \cdot (\Hs \lambda(r)) + \ONE & (\D \lambda(r))\transp \\
(\D \lambda(r)) & 0
\end{pmatrix}
\begin{pmatrix}
\dot r \\ \dot d
\end{pmatrix}
= \begin{pmatrix}
0 \\ \sign{s}
\end{pmatrix}.
\end{align}
We integrate this ODE system from $t=s$ to $t=0$ and obtain the orthogonal projection $q=r(0)$ onto $\lambda=0$ surface.

There is, however, one problem with this: The matrix becomes singular as a focal surface of the $\lambda=t$ surface approaches $q$. If the ODE solver does not adapt its step size, this leads to inaccurate results, whereas if it does adapt, the step size may approach zero and the solver gets stuck or an attempt to invert a singular matrix is made. The solution is to limit the step size from below and use \cref{orthoODE} as the predictor in a kind of predictor-corrector algorithm. We found that a good choice for the corrector step is to replace $r(t)$ with 
$\argmin_{\lambda(\tilde r)=t\cdot \sign s} (\abs{\tilde r-v})$, found using projected gradient descent initialized with $r(t)$. Afterwards, $d$ needs to be updated as well: $d \eqdef \tilde d = \frac{D\lambda (r)\cdot(v - \tilde r)}{\abs{D\lambda (r)}^2}$. With a high-order adaptive step size control scheme (We used Dormand-Prince), the corrector usually has nothing to do because the predicted point already is very accurate. Only when the step size limitation becomes relevant does it have to do a few iterations.

\subsection{Tracing the Orthogonal Projection of a Curve}
Once we have found the orthogonal projection of a point $v$, we can compute its derivatives with respect to changes in $v$, and hence we can trace the orthogonal projection of a differentiable curve. So, let $v$ and $q$ now again be curves parameterized by curve parameter $t$ (unrelated to the use of $t$ in the previous paragraph), and let $d$ be a real-valued function of $t$ so that $v = q + d N_V(q)$, where $N_V(q)$ is the unit normal to the SSB in voltage space.
\footnote{We  tested the 
respective tracing method also for the SSB in power space cf. \cite{gruhl,hein} ; here we 
report  details  only on  the  voltage  space case.}
Then $d = (v - q)\transp N_V$, and by differentiating these with respect to $t$, we get
\begin{align} 
\label{orthoDotd}
\dot v - \dot q & = \dot d  \cdot N_V(q) + d \cdot (\D N_V) \dot q  \\
\dot d & =   ( v - \dot q)\transp N_V(q) +  (v - q)\transp (\D N_V(q)) \dot q .
\end{align}
Here, $\dot q$ is orthogonal to $N(q)$ and $(v - q)$ is parallel to $N_V(q)$ and thus orthogonal to the image of $\D N_V(q)$, which is the tangent plane:
\begin{align} 
\dot d & = \dot v \transp N_V(q),
\end{align}
hence \eqref{orthoDotd} becomes the following linear equation system for $\dot q$, given $\dot v$:
\begin{align} 
(d \cdot (\D N_V(q)) + \ONE) \dot q & = \dot v  - (\dot v \transp N_V(q)) N_V(q).
\end{align}
\eqref{Lij} tells how $\D N = - W$ can be computed.

\begin{figure}[tp]
\fbox{
	\parbox{8.5cm}{
		\center
		\includegraphics[width=8cm]{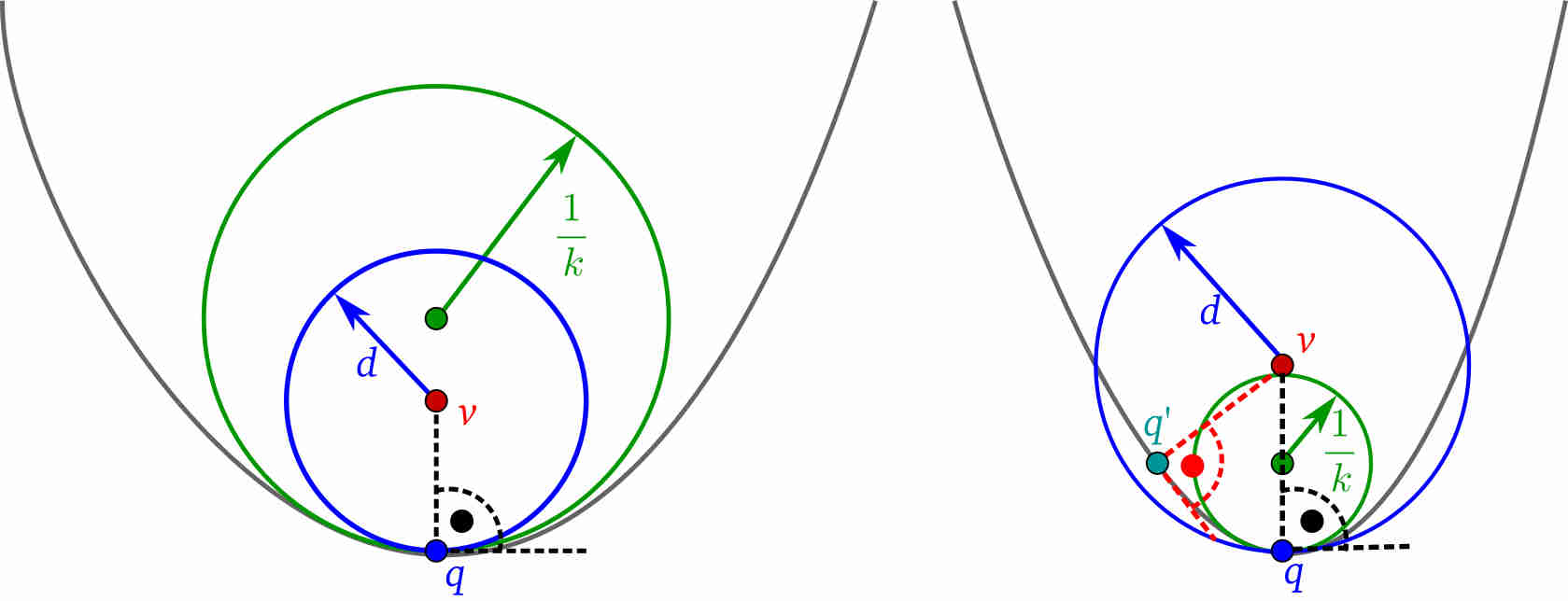}
		\caption{Illustration of necessary condition for orthogonal projection $q$ of $v$ to be the closest point: If distance $d$ is larger than the radius of curvature $1/k$ on the right, there is always another point like $q^\prime$ closer to $v$ than $q$ is. \skipall{Here, $q^\prime$ is also an orthogonal projection and a globally closest point, but already points infinitesimally adjacent to $q$ are closer to $v$ than $q$ is, because in second order they follow the osculating green circle of radius $\frac{1}{k}$, each of whose points except $q$ lies inside the blue circle of radius $d$ around $v$.}
		 }
		 \label{figKruku}
	}
}
\end{figure}

The function $d$ tells the distance between $v$ and $q$. When it is smaller than the smallest positive\footnote{``Positive'' means curving towards $v$; We explain here only the case where $d\leq 0$ and therefore the normal points away from $v$.} radius of curvature of the SSB at $q$, $q$ is a point on the SSB guaranteed to be locally closest to $v$. The smallest positive radius of curvature is the reciprocal of the largest eigenvalue of the Weingarten map. 
If $d$ is larger than the smallest radius of curvature, then $q$ cannot possibly be the closest point to $v$ on the SSB. See \eqref{figKruku} for an illustration, adapted from the master thesis of Gruhl \cite{gruhl}.
\subsection{Local  topology of SSB }
We focus on discussing the local  SSB topology in voltage space. Often  computations  are implicitly restricted to analyzing  restrictions of the  power flow map $F : V \rightarrow P$  practically  to a local neighborhood  
$U(v_0)$  of  a point  $v_0$  on the SSB in voltage space, assuming that $U(v_0)$ is
a topological  unit half disc with boundary  on the SSB,  thus topologically
equivalent to  the bordered  half-space of  $\mathbb{R}^n$  with euclidean norm  $ |v |  <  1$ 
and coordinate  $v^n$ of  $v$ being positive. 

Indeed for practical computing  mostly only  solutions of the power flow 
map  to its restriction being the  homeomorphism  
\begin{equation} \label{eq:whatever}
F : U(v_0) \rightarrow F( U(v_0))
\end{equation}
are relevant.  Thus like  in many other  power flow papers this is a silent  
technical assumption we implicitly  generally made in this paper as well as we 
did not want to mess with multiple  possibly irrelevant solutions, excluding them 
by restriction to case \eqref{eq:whatever}.   Without this assumption, the need for
formal  pre-image specification would aggravate precise descriptions  of
computations without improving insights. 

\begin{figure}
	\fbox{
		\parbox{7.5cm}{
			\center
			\includegraphics[width=7cm]{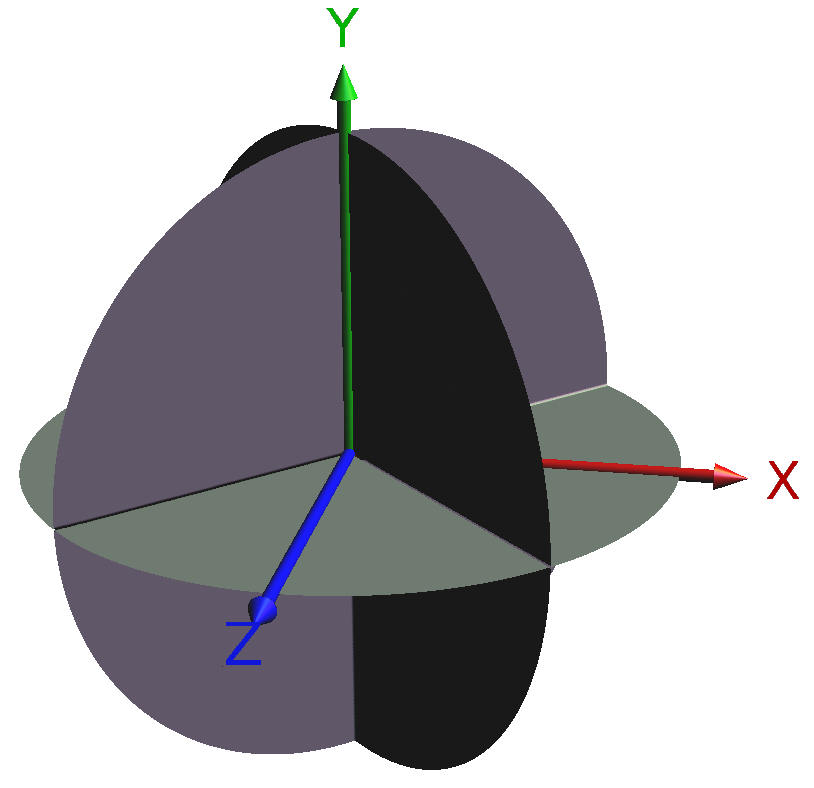}
			\caption{A determinant isosurface with non-manifold lines}
			\label{figDetDegen0} 
		}
	}
\end{figure}

In \cite{wolter2019differential} we proved a practical criterion assuring that the restriction $F$  in  \eqref{eq:whatever} 
would be locally possible as it could be tested that locally close to some point on the SSB, the SSB would be a proper regular rank $n-1$  differential hyper surface iff at $v_0$ the gradient of $\det ( DF(v_0))$ is  nonzero, the latter being equivalent to the easy  test that  the kernel (zero space) of $DF(v_0)$ has dimension one. Beyond that by inspecting low dimensional situations the generically $n-1$ dimensional hyper surface structure of the SSB  even in dimension 3 easily may have partial $n-2$ dimensional submanifolds, cf. fig. \ref{figDetDegen0}.

Beyond the practical relevance for checking necessary conditions for at least 
locally topologically safe conditions for numerically computing  solutions for 
power flow equations we finally would like pointing to an also practically  
relevant insight  disproving  a concrete  erroneous  result in a major 
practically important report \cite{ghys2016singular}. The latter erroneous result got by  \cite{ghys2016singular} highly precise  numerical  spectral  computations was observing fork bifurcations in  planar intersection curves of SSB in voltage space contained in the intersection  set of SSB with a planar family of voltage rays. Those fork type  topological  bifurcations  observed in  \cite{ghys2016singular} cannot occur as the topological structure of the zero set of $\det (F ( v) )$ here restricted on a 2d-plane being a real zero set of an algebraic function in two real variables can only have points with an even number of out going branches, cf. \cite{wolter2019differential, ghys2016singular}.

\begin{figure}
	\fbox{
		\parbox{7.5cm}{
			\center
			\includegraphics[width=7.4cm]{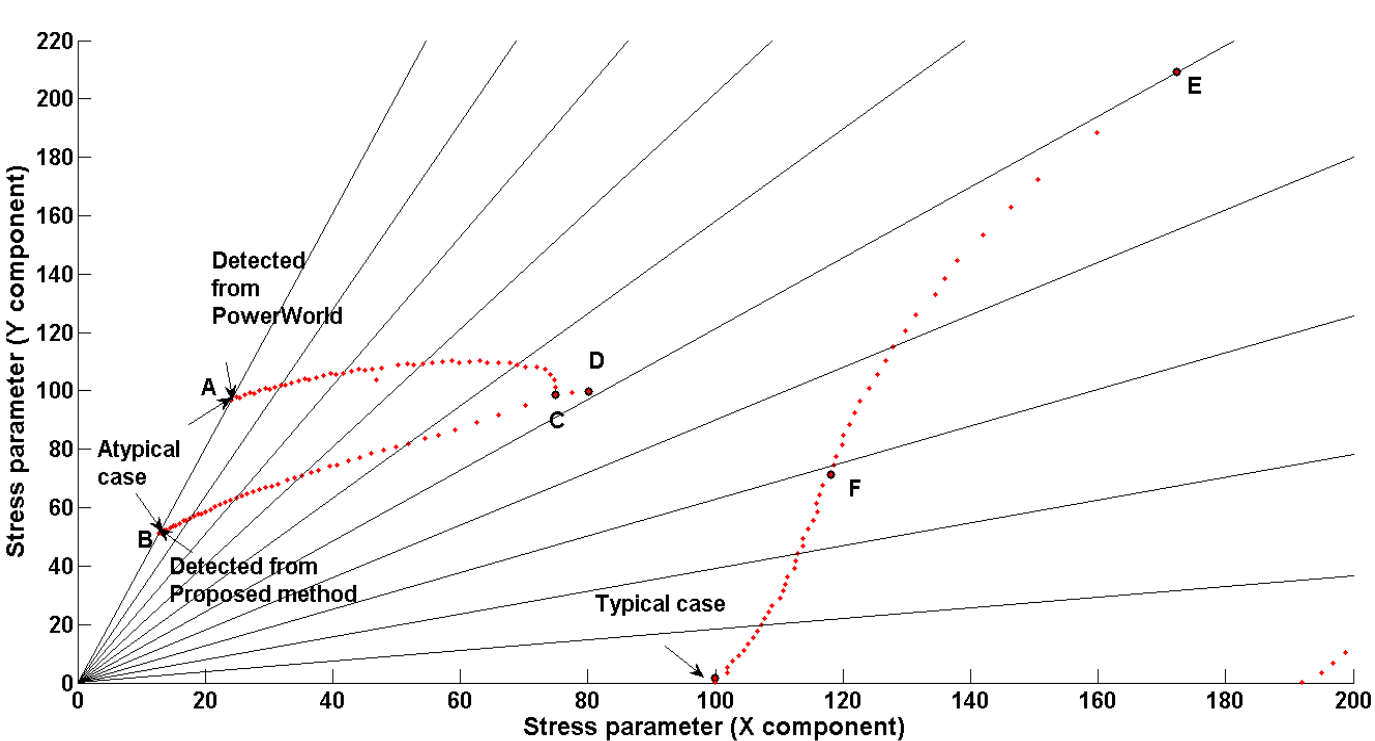}
			\caption{Figure 26 in \cite{makarov2014non}. Reproduction of the graphic should be fair use both because it is for purposes of criticism and because this is scholary work, see 17 U.S. Code § 107.}
			\label{figWrong} 
		}
	}
\end{figure}



\subsection{Geodesic coordinates for solution sets}

We consider the potential advantage of systematic and comprehensive numerical parameterizations using geodesic coordinates for all types of constrained solution sets of power flow equations being likely the most important advantage for using computational differential geometry in power flow computing. For detailed descriptions we refer to \cite{wolter2019differential}, and present here only figures thereby briefly indicating options for using respective numerical geodetic coordinates. Connected components of solution sets of nonlinear equation systems underlying non-linear algebraic constraints even if they generically define regular submanifolds or only 2d-surfaces in high-dimensional Euclidean space are generally hard to endow with coordinates.
We view these coordinates as numerical land charts supporting systematic efforts for registering and accessing in principle all solution points in the connected component of the submanifold partial to the algebraic solution set. Here our computational engineering seemingly still new concept employing numerical geodetic coordinates generally appears being the only generic way providing systematic approaches towards the aforementioned goals. This is so partly the validity of the theorem of Hopf and Rinow: Saying in the situation that any point in the connected component of the solution set can be reached from any other point in the component by a (shortest) geodesic line contained in the component if that geodesic starts with the right initial direction. After fixing the initial direction the 1d-path of the geodesic is uniquely defined. This yields in principle a method for systematically reaching and covering all points in the connected component of the solution set and thus comprehensively and systematically parameterizing all the component of the solution set. Even more always locally in a neighborhood of the start point and under some conditions globally this parametrization, eg., by geodesic polar coordinates is a diffeomorphism. Beyond that we can compute and use the Jacobian of those geodesic coordinates to trace a geodesic connecting path back into the euclidean parametrising coordinate space.


\begin{figure}
	\fbox{
		\parbox{7.5cm}{
			\center
			\subfigure[A geodesic coordinate net on the intersection of the unit sphere and a 3-dimensional SSB in 4-dimensional space, stereographically projected into 3-dimensional space.]{\includegraphics[width=7cm]{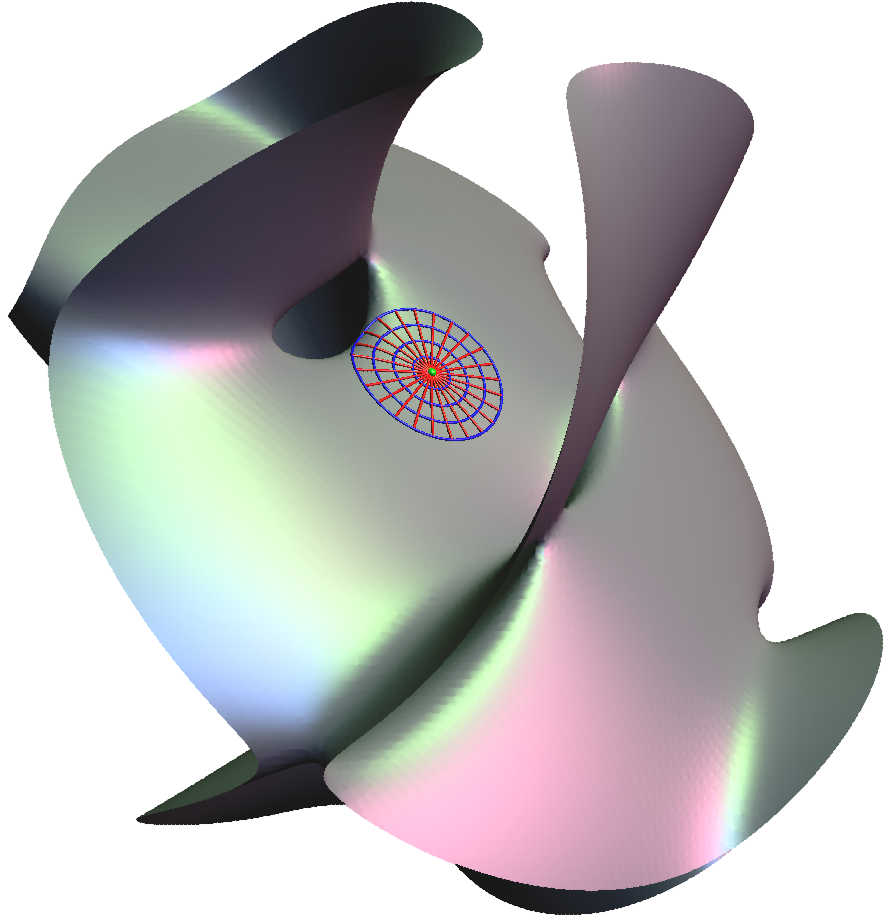}}\\
			\subfigure[The projection of the geodesic coordinate net onto the subspace spanned by coordinate directions $P_1$ and $P_2$]{\includegraphics[width=3.4cm]{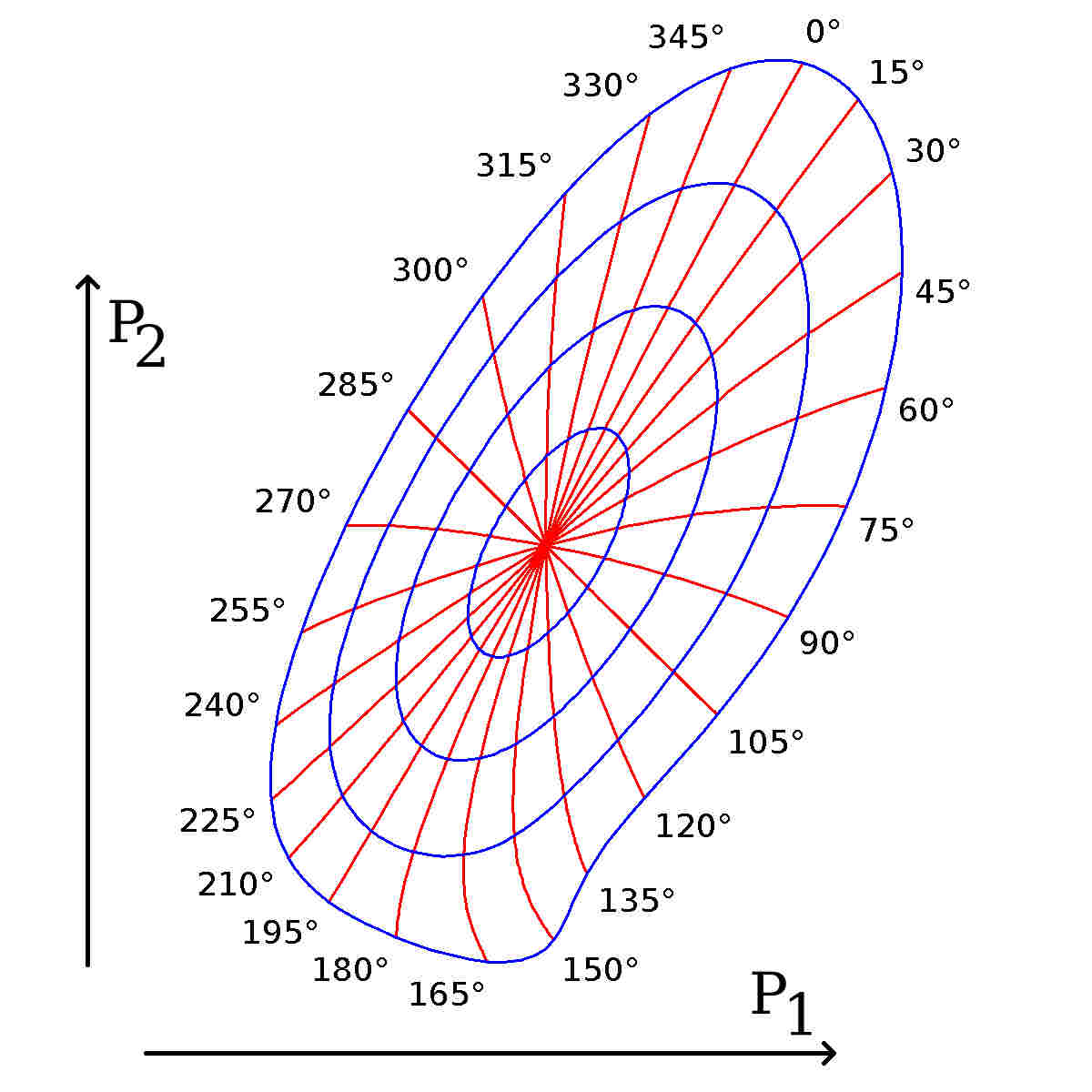}}\quad
			\subfigure[Ditto for coordinate directions $P_3$ and $P_4$]{\includegraphics[width=3.4cm]{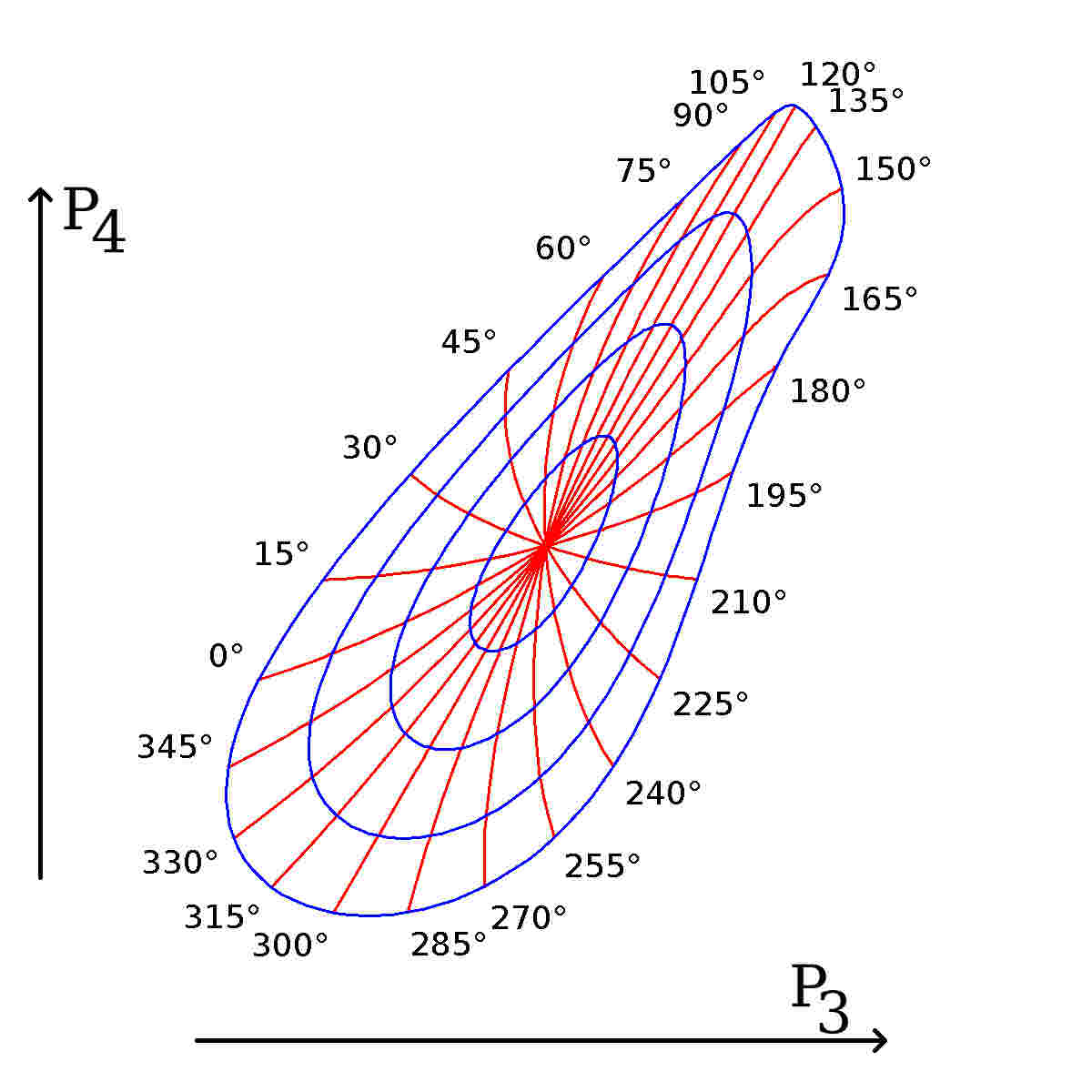}}
			\caption{}
			\label{figGeoProj} 
		}
	}
\end{figure}

In our power flow setting, we have the advantage that the solutions set of the involved algebraic equations (subject to constraints) typically relate to sparse quadratic equations. This makes it computationally feasible  computing families of geodesics contained in and used to parametrize complicated high dimensional submanifolds to the extent being needed despite the
constraints that may nastily constrain them to submanifolds of the SSB.
Here we only indicate the above concept descriptively via figure 4 graphically illustrating a sample case referring to a special situation in power flow computing. We have here a four-dimensional power flow system say defined by a map F from 4d voltage space to 4d power space,  where we have 
two passive and two active power coordinates, their technical meaning being
exchangeable in the context of our computing. Now, in this case, the solution space boundary is a 3d-hyper surface in both voltage and power space. In the context at hand a 2d-manifold in the 3d-SSB has been defined by the additional constraint that it must be contained in a 3d-euclidean unit sphere in voltage space, thus points with Euclidean norm  $|v| = 1$; (it could have been any other
constraint as well. This one was graphically convenient as it allowed projecting this unit sphere nicely stereographically into euclidean 3d-Space.) Now we immediately numerically can match between say the two active and the two passive power parameters employing the underlying geodesic polar coordinates inducing respective 2d- polar coordinates  on the (grey) 2d- manifold in voltage space as well as on the two different respective 2d- planes in power space described
by $(p_1, p_2)$ and $(p_3,p_4)$ respectively. 
This would provide transformations based
on geodesic coordinates, i.e, answering in this situation, (with all points partial to  2D-surface in the 3D-SSB) what would for (given $p_1, p_2$ parameters) be the missing matching $(p_3, p_4)$ parameters, needed to complete the coordinates assuring the point in 4D would still comply with all given constraints. This prototype concept can be extended to far more complicated higher dimensional
scenarios.

\section{Conclusion and Further Work}
We presented various formulas and algorithms for calculating geometric entities associated with the SSB, such as curvatures and orthogonal projections of points and curves. We laid special focus on the SSB in voltage space, which historically has received less attention and which is less accessible computationally. Our algorithm for local inversion of the power flow map allows us to continue a solution to the power flow equations along a curve with high precision. We omitted deeper results on the local topological
structure of SSB, algorithms for computing geodesics and
Jacobeans of geodesic coordinates on SSB, and algorithms for improving the results of optimal power flow computing as to assure certain security constraints a posteriori. We plan to publish these results regularly, for now, they are accessible in \cite{wolter2019differential}.


\subsection*{Acknowledgements}
This work had its origin during a research term of F.-E. Wolter at MIT in summer 2016, supported by a MISTI (MIT Germany) research grant that he initiated together with T. Sapsis, N. Patrikalakis and S. Karaman. This group was later joined by K. Turitsin. During this research term, F.-E. Wolter came into contact with electrical power researchers from Argonne Laboratory (D. Molzahn) and the University of Michigan (Ian Hiskens), who later invited him to visit \cite{wolter2016MIT}. 

All these scientists have been supporting the advancement of the ideas presented in this paper via various discussions and references, which had been both enlightening and encouraging. During visits to the IMI, NTU, which included seminars there \cite{wolter2018NTU,wolter2018CGI}, work done in this project evolved further. Most recently, F.-E. Wolter, during his 2019 visit in Singapore \cite{wolter2019NTU}, had the chance to come into contact with power flow researcher H. D. Nguyen who provided stimulating questions. We thank  Hannes Thielhelm for many stimulating discussion on geodesics and Riemannian geometry closely related to topics in this paper. Thank IMI for the opportunity to present ideas of this project in interdisciplinary seminars and for many stimulating informal discussions with various researchers there, as well as of course for the generous invitations by N. Thalmann that helped building new promising cooperations in Singapore as well as supporting presenting the results to the computer graphics community \cite{wolter2018CGI}.

\nocite{thielhelm2015geodesic,gutschke2015differential,wolter2018CGI,yao2018improving}


\bibliographystyle{IEEEtran}
\bibliography{Paper}
\end{document}